\documentclass[fleqn,usenatbib,useAMS]{mnras}

\usepackage{mathptmx}

\usepackage[T1]{fontenc}

\usepackage{graphicx}	
\usepackage{amsmath}	
\usepackage{amssymb}	

\newcommand{\kev}{keV}

\newcommand{\xmm}{\textit{XMM-Newton}}

\newcommand{\fe}{Fe~K$\alpha$}

\newcommand{\mrk}{Mrk~335}
\newcommand{\oneh}{1H~0707-495}

\title[Warm Coronae in AGNs]{Examining the Physical
Conditions of a Warm Corona in Active Galactic Nuclei Accretion Discs}

\author[D. R. Ballantyne]{
D. R. Ballantyne\thanks{E-mail: david.ballantyne@physics.gatech.edu}
\\
Center for Relativistic Astrophysics, School of Physics, Georgia
  Institute of Technology, 837 State Street, Atlanta, GA 30332-0430, USA\\
}

\date{Accepted XXX. Received YYY; in original form ZZZ}

\pubyear{2019}

\begin{document}
\label{firstpage}
\pagerange{\pageref{firstpage}--\pageref{lastpage}}
\maketitle

\begin{abstract}
A warm corona at the surface of an accretion disc has been proposed as
a potential location for producing the soft excess commonly observed
in the X-ray spectra of active galactic nuclei (AGNs). In order to fit
the observed data the gas must be at temperatures of $\sim 1$~\kev\
and have an optical depth of $\tau_{\mathrm{T}}\approx
10$--$20$. We present one-dimensional calculations of the physical conditions and
emitted spectra of a $\tau_{\mathrm{T}}=10$ or $20$ gas layer subject to
illumination from an X-ray power-law (from above), a blackbody (from
below) and a variable amount of internal heating. The models show that
a warm corona with
$kT \sim 1$~\kev\ can develop, producing a strong Comptonized soft
excess, but only if the internal heating flux is within a
relatively narrow range. Similarly, if the gas density of the layer is too large
then efficient cooling will stop a warm corona from forming. The radiation from the hard X-ray
power-law is crucial in producing a warm corona, indicating that a
warm and hot corona may co-exist in AGN accretion
discs, and their combined effect leads to the observed soft
excess. Intense heating of a warm
corona leads to steep X-ray spectra with ionised
\fe\ lines, similar to those seen in some narrow-line Seyfert 1 galaxies. 
\end{abstract}




\section{Introduction}
\label{sect:intro}
The X-ray spectra of active galactic nuclei (AGNs) at energies $\ga 2$~\kev\ are often well described by
a cutoff power-law continuum \citep[e.g.][]{panessa08,malizia14,ricci17,ricci18,molina19} that is consistent with being
produced by Comptonization of ultraviolet (UV) disc photons in a hot,
tenuous corona
\citep[e.g.,][]{galeev79,hm91,hm93,jiang14,jiang19}. However, below
this energy the majority of AGNs ($\ga 50$\%; e.g.,
\citealt{scott12,winter12,ricci17}) exhibit a significant amount of emission above what is
predicted by extrapolating the power-law to lower-energies. This `soft
excess' has been observed in many AGNs for decades \citep[e.g.,][]{arnaud85,pounds86,wf93,leighly99}, but the
origin of this emission is not understood. Observationally,
the soft excess appears as a featureless, smoothly varying continuum
that can be described by thermal emission with temperatures $\sim 0.1$--$0.2$
keV \citep[e.g.,][]{bianchi09}. The blackbody temperatures do not show any correlation
with the AGN black hole masses or accretion rates, indicating that the
emission is not simply thermal radiation from the hottest part of the
accretion disc \citep[e.g.,][]{gd04,bianchi09}.

The lack of variation observed in the temperatures has led to models
where the soft excess is strongly influenced by atomic processes \citep[e.g.,][]{gd04}. For
example, reprocessing of the X-ray
power-law in the inner accretion disc (i.e., X-ray
`reflection'; \citealt{fr10}) naturally predicts a strong bremsstrahlung spectrum at
energies $\la 1$~\kev\ that may be accompanied by a forest of spectral
lines depending on the ionization state of the reflecting region
\citep[e.g.,][]{rf93,brf01,garcia10}. If this reflection spectrum is produced close to a rapidly
spinning black hole, light-bending effects will combine with strong relativistic blurring of the spectrum to
produce a featureless soft excess that can describe the observations
in some cases \citep[e.g.,][]{crummy06,walton13,garcia19}. However,
the significant relativistic blurring needed by the model to explain
the featureless soft excess often leads to solutions with extremely rapidly
spinning black holes and very low coronal heights, even with models
that allow for large disc densities \citep[e.g.,][]{garcia19,jjiang19}. While it is clear that soft emission from the disc
reflection spectrum must contribute to the soft excess in many
AGNs, the fact that models are pushed into an extreme corner of
parameter space is concerning. Spectral modeling that allows for an
extra emission component to help with the soft excess fit does not
require such extreme values for the black hole and corona properties
\citep[e.g.,][]{kb16}. Therefore, it is worth considering how this
extra emission component will interact with the disc reflection
physics. 

An alternative model for the soft excess returns to the idea of direct
emission from the AGN accretion disc, but now proposes that the
thermal disc emission passes through a thick ($\tau_{\mathrm{T}} \sim
10$--$20$, where $\tau_{\mathrm{T}}$ is the optical depth to Thomson scattering),
warm ($kT \sim 0.1$--$1$~keV) Comptonizing layer \citep[e.g.,][]{mag98,czerny03,petrucci13,petrucci18,kd18}. If this layer is
common in AGNs, then the thermal disc emission will be broadened up to
soft X-ray energies, and, depending on how the warm layer is heated, may appear insensitive to changes in the black hole
mass and accretion rate. This `warm corona' would need to reside at or
near the surface of the accretion disc, and could be heated by
accretion energy dissipated within the disc and transported vertically
(perhaps by magnetic fields; e.g., \citealt{jiang19}) to the disc surface. \citet{petrucci18} used a
phenomenological warm corona model to successfully fit nearly 100
optical-to-X-ray spectra of 22 Seyfert galaxies observed with
\xmm. However, it is unclear if this scenario is physically
plausible. For example, \cite{roz15} considered the hydrostatic
balance of a warm corona on top of a cool accretion disc and found
that balance was only achieved for $\tau_{\mathrm{T}} < 5$ without additional
pressure support from, e.g., magnetic fields. More recently, \citet{garcia19} showed that gas at the
temperatures and optical depths expected in a warm corona will retain significant photoelectric
opacity and therefore any radiation passing through the gas will be
imprinted with many soft X-ray absorption lines which are not observed
in AGN data.

Despite these difficulties, the substantial uncertainty in our
understanding in how accretion energy is transported and dissipated through a disc-corona
system means that it is important to continue to closely examine the warm corona scenario. In this paper, the atomic and radiative processes within a
warm corona are treated in detail in order to accurately predict its
radiative signatures, and compare them against those needed in the
phenomenological models \citep[e.g.,][]{petrucci18}. In particular, we
note that the warm corona scenario still requires the presence of the hard X-ray
power-law and its associated reflection spectrum. In fact, the X-ray
power-law will be illuminating the surface of any warm corona,
significantly impacting the physical conditions within that layer. It
is important, therefore, to consider the effect of the power-law
heating on the physical properties of a warm corona and to determine
if this situation can still provide an explanation for the soft excess
in AGNs. This paper investigates this question by computing the
physical conditions of a layer of gas subject to (a)
uniform heating throughout by an assumed accretion flux, (b) an
external hard X-ray power-law, and (c) a soft blackbody injected from
below. These calculations will determine under which conditions, if
any, a warm corona with properties similar to those inferred from
observations can exist when including the effects of the power-law
irradiation. The next section describes the set-up of the calculation,
with the results presented in Sect.~\ref{sect:results} and discussed
in Sect.~\ref{sect:discuss}. The last
section (Sect.~\ref{sect:concl}) summarizes the key findings and
conclusions.   

\section{Calculations}
\label{sect:calc}
The calculations are based on the one-dimensional, time-independent reflection models of
\citet{brf01,brf02} in which the surface of the accretion disc,
treated as a slab of gas with hydrogen number
density $n_{\mathrm{H}}$ and Thomson depth $\tau_{\mathrm{T}}$, is
irradiated by an external X-ray power-law. A blackbody spectrum can also be introduced into the lowest zone of the
slab to account for the emission from the bulk of the accretion disc
laying below the slab. As described in the above articles and
references therein, the temperature structure of the slab is solved by
considering multiple heating and cooling processes in the gas, and the
spectrum emitted by the surface (comprising the sum of the reprocessed
power-law radiation, transfered blackbody flux, and diffuse emission
from the gas) is computed once the slab
reaches thermal balance. The temperature of AGN accretion discs is
expected to be large enough that hydrogen is completely ionized \citep{ss73}
filling the slab with free electrons. Thus, the calculation carefully
treats the effects of Compton scattering which is crucial in
redistributing photon energies and shaping the emitted spectrum
\citep{ross79,rf93}. This fact, combined with the broad energy range
treated by the calculations ($0.8$~eV to $98$~\kev), allows for the
simultaneous prediction of both the soft-excess and hard X-ray
reflection properties produced by an irradiated disc surface.
  
In order to self-consistently predict the spectrum emitted by an
optically-thick warm corona in the presence of a hard X-ray power-law,
we consider the total emergent flux emitted by an accretion disc (per disc side) at radius $R$, $D(R)$. Following the warm corona hypothesis, a fraction
$h_f$ of this flux is assumed to be uniformly dissipated as heat throughout the
irradiated slab. This leads to an additional heating function
(in erg cm$^3$~s$^{-1}$) in the calculation,
\begin{equation}
  \mathcal{H} = {h_f D(R) \sigma_{\mathrm{T}} \over \tau_{\mathrm{T}}
    n_{\mathrm{H}}}
  \label{eq:hfunct}
\end{equation}
where $\sigma_{\mathrm{T}}$ is the Thomson cross-section. This heating
function corresponds to a constant heating rate per particle over $\tau_{T}$.  

The hard X-ray power-law that irradiates the disc is assigned a total
flux of $f_{X}D(R)$, following the expectation that the hot corona is
heated by accretion energy transported outside of
the disc \citep[e.g.,][]{jiang14,jiang19}. In order to
consider a large range of $h_f$, we fix $f_X=0.1$, a value similar to
the X-ray bolometric corrections inferred in local AGNs \citep[e.g.,][]{vf07,netzer19}. The illuminating hard X-ray
power-law is defined to have an exponential cutoff at $30$~eV to mimic
its expected origin as Comptonized accretion disc emission
\citep[e.g.,][]{coppi99}. In addition, both the photon-index
of the power-law, $\Gamma$, and its high-energy cutoff energy,
$E_{\mathrm{cut}}$ are fixed at $\Gamma=1.9$ and
$E_{\mathrm{cut}}=220$~keV, respectively. These values are typical of
bright Seyfert galaxies \citep[e.g.,][]{ball14,ricci17,ricci18}. Finally, the
remaining flux, $(1-f_X-h_f)D(R)$, is injected into the bottom of the
slab as a blackbody with a temperature given by the standard blackbody
formula. This setup allows an investigation of the emitted spectrum
produced by the heated layer of an accretion disc while constraining
the total energy available for heating the gas.

The accretion flux $D(R)$ is calculated from the standard thin disc
equation \citep[e.g.][]{ss73,mfr00} with a black hole mass of $2.5\times
10^7$~M$_{\odot}$, an accretion rate of $0.6$ times the Eddington
value, and $R=5$~Schwarzschild radii. These values result in
$D(R)=5.175\times 10^{16}$~erg~cm$^{-2}$~s$^{-1}$. Spectral fitting
results using the warm corona hypothesis find that the heated layer
has $\tau_{\mathrm{T}} \approx 10$--$20$ \citep[e.g.,][]{petrucci18}. In addition, the heating and
cooling processes in the gas (e.g., Compton scattering,
bremsstrahlung) are strong functions of density, so the
spectra produced by the models will depend on the assumed density of the slab
\citep[e.g.,][]{ball04,garcia16,jjiang19}. Therefore, the following
set of four scenarios are considered to explore the properties of a
warm corona: $n_{\mathrm{H}}=2\times 10^{14}$~cm$^{-3}$
($\tau_{\mathrm{T}}=10$, $20$) and $n_{\mathrm{H}}=5\times 10^{14}$~cm$^{-3}$
($\tau_{\mathrm{T}}=10$, $20$). Eight reflection models, from
$h_f=0.1$ to $0.8$, are computed for each ($n_{\mathrm{H}}$,
$\tau_{\mathrm{T}}$) pair. The resulting X-ray emission from and physical
conditions within the model warm coronae are discussed below.

\section{Results}
\label{sect:results}
The goals of the calculations are to (1) test the likelihood of forming a
$kT \sim 1$~keV warm corona when the heated slab is subject to
irradiation from a hard X-ray power-law, and (2) determine if the
resulting emission and reflection spectrum exhibits a strong,
relatively smooth soft excess. The models are evaluated against these
two criteria while also considering the impact of changing the optical
depth and density of the heated surface.

\subsection{Making a Warm Corona: Testing the Effects of Optical
  Depth}
\label{sub:tau}
\begin{figure*}
  \includegraphics[width=0.98\textwidth]{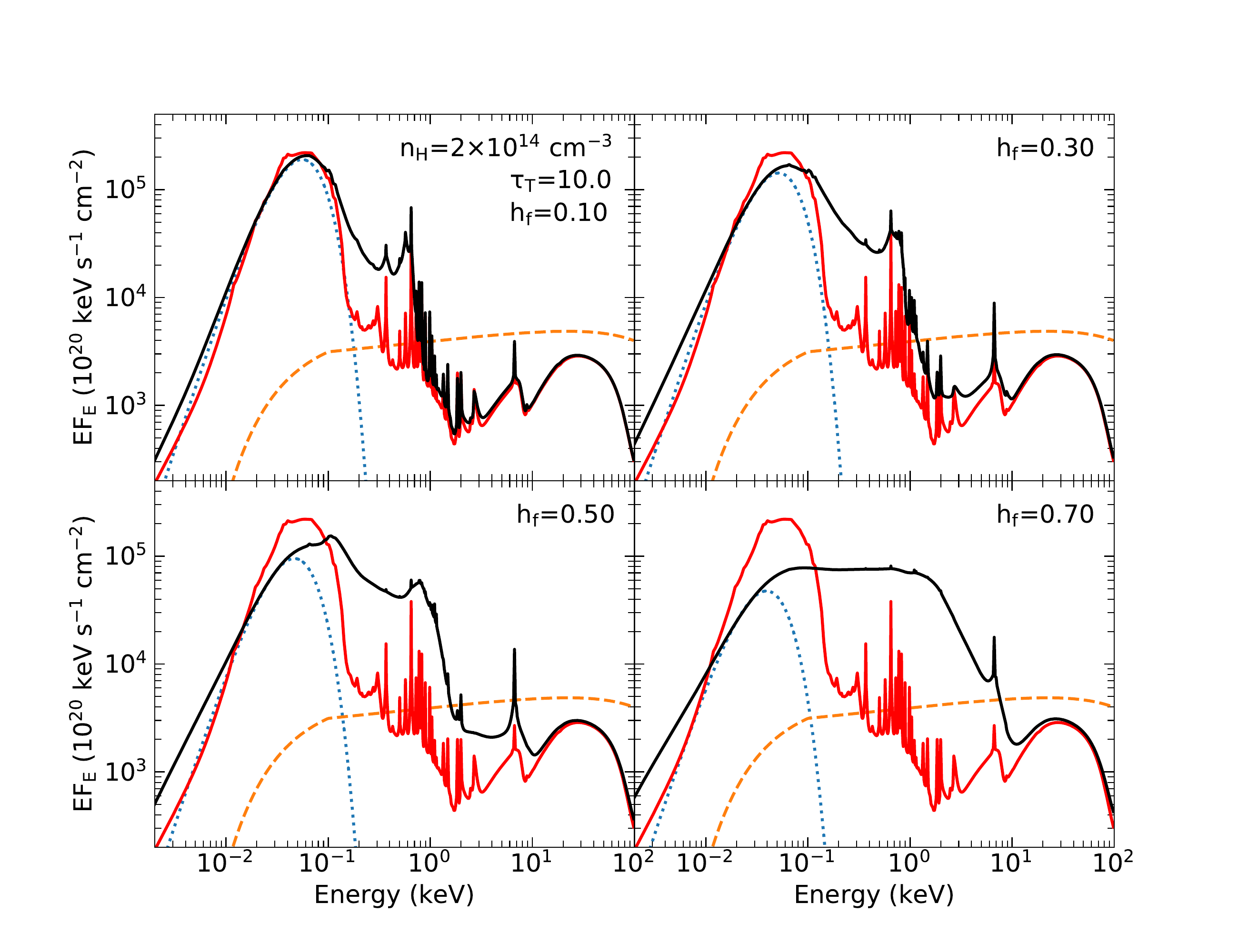}
  \caption{The black curves in each panel show the predicted emission
    and reflection
    spectrum from a constant density slab with
    $n_{\mathrm{H}}=2\times 10^{14}$~cm$^{-3}$ and $\tau_{\mathrm{T}}=10$. The slab is irradiated from above by a fixed cutoff
    power-law spectrum (dashed curves) and below by a blackbody (dotted
    curves). The different panels show the effects of increasing the
    fraction $h_f$ of the accretion flux that is injected as
    heat into the slab (see Sect.~\ref{sect:calc}). The solid red curve
    in each panel plots the reflection spectrum for the null
    hypothesis model with \emph{no} additional heating. In all cases,
    the power-law, blackbody, and heat fluxes sum to
    $D(R)$. The figure clearly shows that releasing heat into the surface of
    the slab strongly affects the shape of the hard X-ray reflection
    spectrum, including producing a strong soft excess (e.g., when
    $h_f=0.50$). \label{fig:nh2e14tau10spect}}
\end{figure*}

The phenomenological fits by \citet{petrucci18} found that the warm
corona model for the soft excess required that the warm layer have
$\tau_{\mathrm{T}} \approx
10$--$20$. The black solid lines in Figure~\ref{fig:nh2e14tau10spect} shows the results of four
of our reflection calculations for the case of
$n_{\mathrm{H}}=2\times 10^{14}$~cm$^{-3}$ and
$\tau_{\mathrm{T}}=10$. The four panels of the figure display the
impact of increasing the heating fraction $h_f$ from $0.1$ to $0.7$.
Each panel also illustrates the radiation fields impacting the model
slab, with the external cutoff power-law irradiating the surface of
the layer shown as the dashed line, and the blackbody injected at the
bottom shown as the dotted line. Lastly, to illustrate the impact of
the heating injected into the slab, the solid red curve in each panel shows the
emission and reflection spectrum of a model with no artificial
heating. Every model, including the one with no heating, is subject to
the same total energy flux of $D(R)$ ensuring that the only difference
between models is how much energy is injected as heat using the
$\mathcal{H}$ function (Eq.~\ref{eq:hfunct}).

Comparing the black and red curves clearly shows the impact of the
heat released into the model slab. As the power-law is constant in each model, the spectrum changes as a
function of $h_f$ simply due to the response of the additional heat
injected throughout the layer. This heat leads to two significant
effects. First, it results in enhanced Compton up-scattering of the
blackbody throughout the layer, and, second, it increases the overall
ionization state of metals in the gas. Even at the low level of
$h_f=0.10$, Compton up-scattering of the blackbody creates an excess
of emission between $0.1$ and $\approx 0.7$~\kev, although the
temperature of the slab ($kT \la 0.2$~keV) is low enough that
significant recombination line emission is imprinted on the
spectrum, and the hard X-ray spectrum is nearly unaltered. At this
small level, the gas and resulting spectral properties are still
dominated by the effects of the power-law.

The impact of the heating is seen more clearly in the $h_f=0.3$ model,
where a stronger soft excess is now apparent, along with reduced line
emission below $\sim 1$~\kev. Crucially, the \fe\ line is now
significantly stronger than in the model with no heating. In both
cases it originates from recombination onto He-like Fe, but the extra
heat in the $h_f=0.3$ model has increased the population of highly
ionized iron in the gas, leading to a much stronger line.

\begin{figure}
  \includegraphics[width=0.47\textwidth]{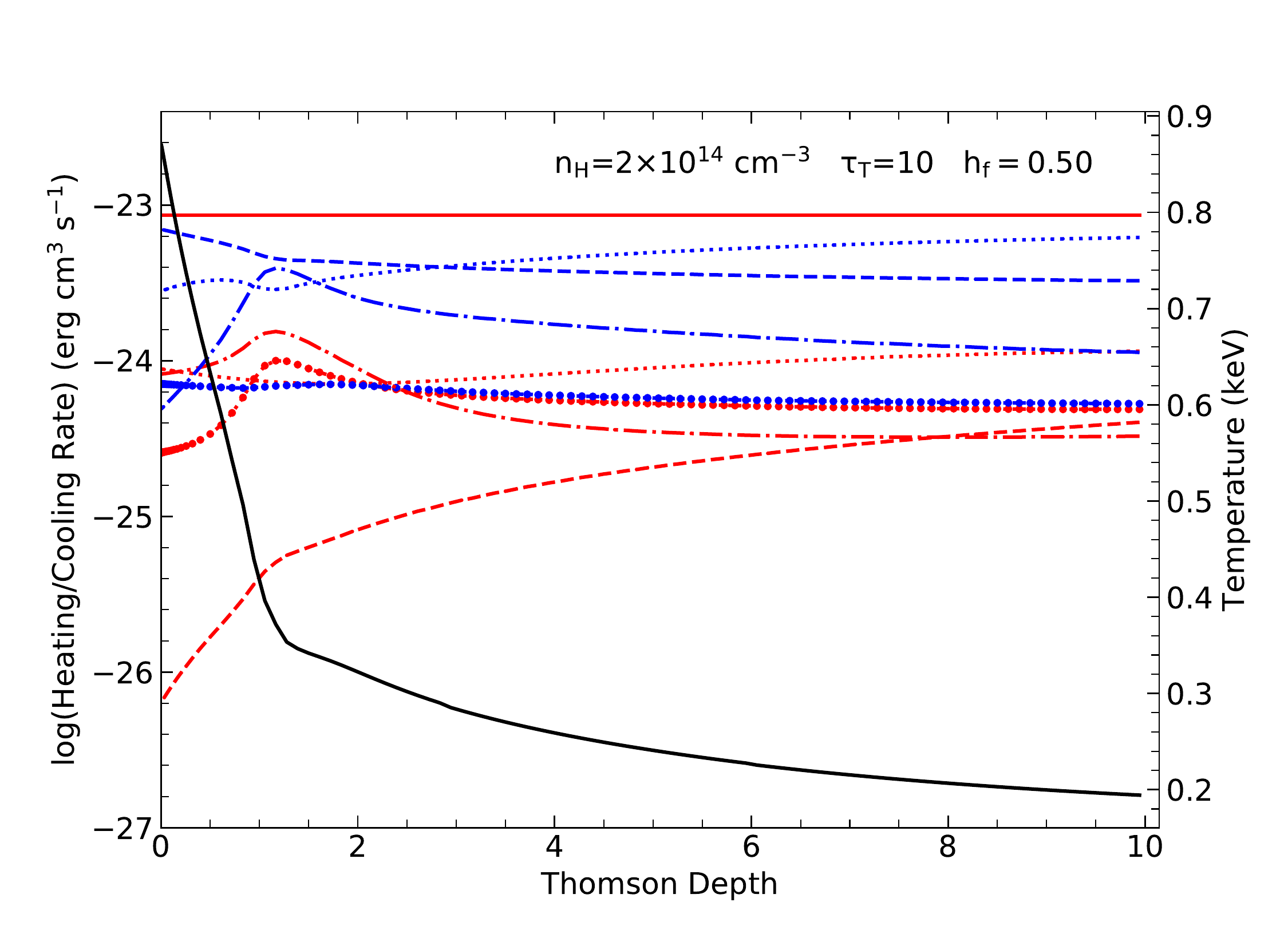}
  \caption{A close look at the heating and cooling processes (red and
    blue lines, respectively) in the
    heated slab with $n_{\mathrm{H}}=2\times 10^{14}$~cm$^{-3}$,
    $\tau_{\mathrm{T}}=10$, and $h_f=0.5$. The different processes are
    denoted by different line styles: Compton heating and cooling
    (dotted lines), bremsstrahlung heating and cooling (short-dashed
    lines), photo-ionization heating and line cooling (dot-dashed
    lines), and recombination heating and cooling (thick dotted
    lines). The solid horizontal line indicates the heating injected
    into the layer (Eq.~\ref{eq:hfunct}) which is constant in this
    constant density slab. Finally the solid black line and right-hand
    axis shows the temperature of the slab, indicating that it has an
    appropriate temperature for a `warm corona'. \label{fig:nh2e14rates}}
\end{figure}

Both of these effects are further enhanced in the $h_f=0.50$ model,
where the reflection spectrum now clearly exhibits a strong, and
relatively featureless, soft excess below $\sim
2$~\kev. To better understand the physics leading to this spectrum,
Figure~\ref{fig:nh2e14rates} plots the heating and cooling rates for
all the processes considered by the model as a function of
$\tau_{\mathrm{T}}$.
The solid horizontal line at the top of the plot shows that $\mathcal{H}$
(Eq.~\ref{eq:hfunct}), the coronal heating injected into the layer,
dominates the heating of the gas at all depths, even close to the
surface. Cooling of the gas is dominated by Compton scattering for
$\tau_{\mathrm{T}} \ga 3$, and by bremsstrahlung at lower Thomson
depths. Indeed, as seen below, free-free cooling, which increases as
$T^{1/2}$ \citep[e.g.,][]{rl79}, becomes increasingly important in shaping the spectrum
as $h_{f}$ increases. Fig.~\ref{fig:nh2e14rates} also shows the
temperature of the slab (solid black line and right-hand axis), which
varies from $\approx 0.2$~\kev\ at the bottom to $\approx 0.85$~\kev\
at the surface. Therefore, the temperature of this $\tau_{\mathrm{T}}=10$
slab is consistent with the properties of a warm corona inferred by
spectral modeling \citep[e.g.,][]{petrucci18}, and also appears to
produce a strong and relatively featureless soft excess.

The reflection and emission spectrum of the slab changes dramatically
in the $h_f=0.70$ model. In this case, the gas is heated so much by
the coronal heating that bremsstrahlung and Compton cooling dominate
the other cooling processes by approximately an order of magnitude,
and the emission spectrum is dominated by a Comptonized bremsstrahlung
continuum with an \fe\ line superimposed on the rapidly declining tail
of the spectrum. The external hard X-ray power-law only influences the
spectrum at energies $\ga 10$~\kev. The temperature of the slab varies
from $0.4$~\kev\ to $1.4$~\kev, so it is not particularly different
from the $h_f=0.5$ model. However, the cooling processes in the gas
have sufficient temperature sensitivity that relatively modest
increases in the gas temperature will significantly alter the emission
spectrum. Clearly, there is a maximum amount
of coronal heating that can produce a warm corona with the correct
properties and yield an appropriate spectral shape. 

\begin{figure*}
  \includegraphics[width=0.98\textwidth]{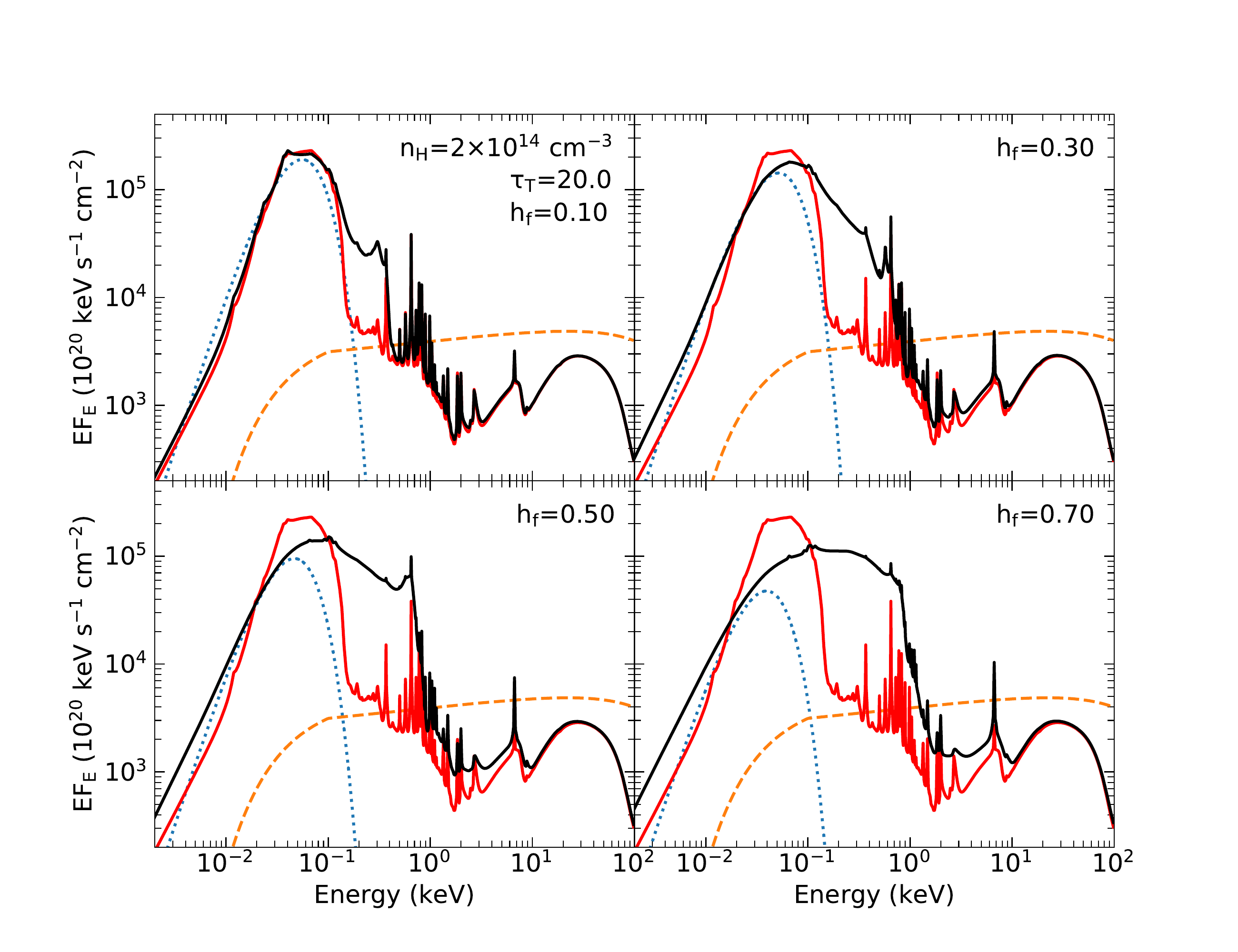}
  \caption{As in Fig.~\ref{fig:nh2e14tau10spect}, but now for a
    thicker $\tau_{\mathrm{T}}=20$ layer. As the same heating flux is
    spread over twice the optical depth, the gas is not heated to the
    same extent, and it requires a larger $h_f$ to obtain a spectrum
    with a strong soft excess. \label{fig:nh2e14tau20spect}}
\end{figure*}

\begin{figure}
  \includegraphics[width=0.47\textwidth]{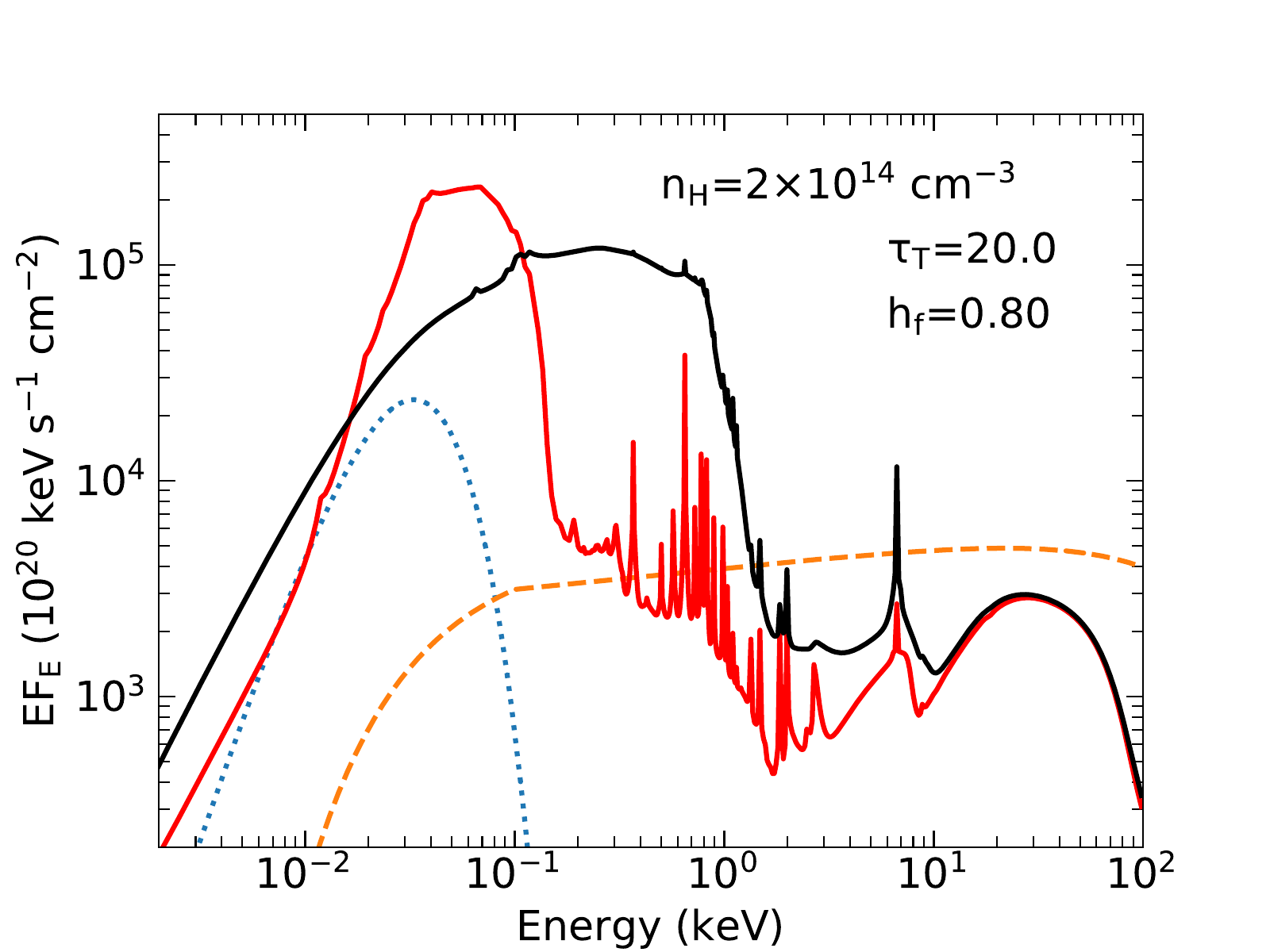}
  \caption{The model results for a $n_{\mathrm{H}}=2\times
    10^{14}$~cm$^{-3}$ slab with $\tau_{\mathrm{T}}=20$ and
    $h_f=0.8$. The line styles distinguish the various
    radiative components as described in the caption to
    Fig.~\ref{fig:nh2e14tau10spect}. Very optically thick warm corona
    can still generate a strong soft excess, but requires that a
    large fraction of the accretion energy be dissipated in the
    heated layer. \label{fig:nh2e14hf08}}
\end{figure}

The effects of doubling the optical depth of the model slab to
$\tau_{\mathrm{T}}=20$ are shown in Fig.~\ref{fig:nh2e14tau20spect}.
As expected from Eq.~\ref{eq:hfunct}, doubling $\tau_{\mathrm{T}}$
will reduce $\mathcal{H}$ everywhere by a factor of $2$. Therefore,
the model slabs are cooler for each $h_f$ as compared to the
$\tau_{\mathrm{T}}=10$ model, and a larger $h_f$ is needed to produce a
strong and smooth soft excess. Fig.~\ref{fig:nh2e14tau20spect} shows
that only for $h_f \ga 0.7$ will a potential soft excess be formed by
the thick layer. The gas temperature of the $h_f=0.7$ model reaches
$\sim 0.5$~\kev\ at the surface. Thus, while it is more challenging to
form a successful warm corona in the thick $\tau_{\mathrm{T}}=20$
layer, the difficulty in heating the gas also means that it will be
very hard to overheat the gas and produce the Comptonized
bremsstrahlung spectrum seen in the $h_f=0.70$, $\tau_{\mathrm{T}}=10$
model. Indeed, the $h_f=0.8$, $\tau_{\mathrm{T}}=20$ model produces a
spectrum with an acceptable strong and relatively featureless soft
excess with a maximum gas temperature of $\sim 0.6$~\kev\
(Fig.~\ref{fig:nh2e14hf08}).

\subsection{Making a Warm Corona: Testing the Effects of Density}
\label{sub:density}
\begin{figure*}
  \includegraphics[width=0.98\textwidth]{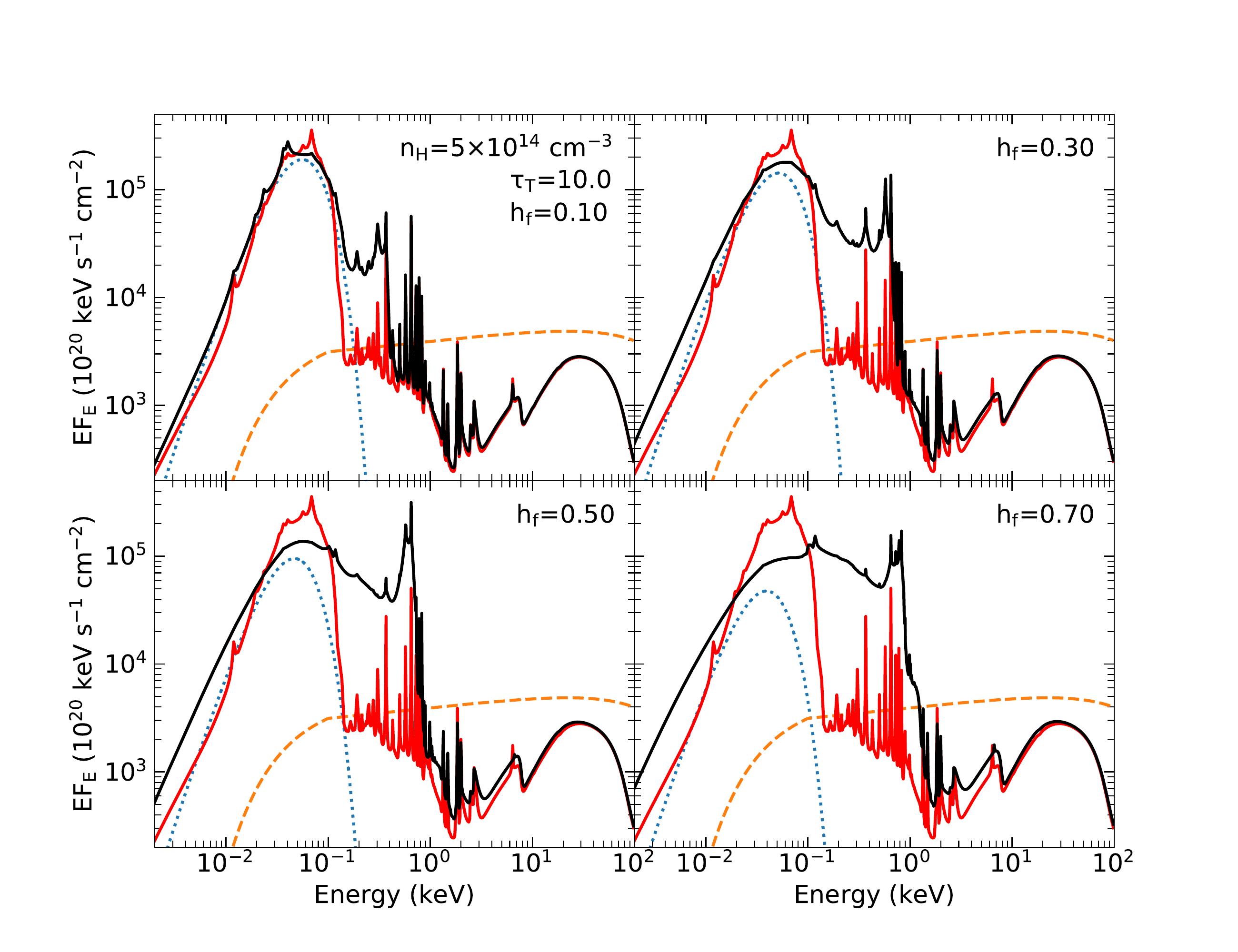}
  \caption{As in Fig.~\ref{fig:nh2e14tau10spect}, but now for a
    denser layer with $n_{\mathrm{H}}=5\times 10^{14}$~cm$^{-3}$. The
    significant increase in density (by 250\%) reduces the overall
    ionization and coronal heating rate (Eq.~\ref{eq:hfunct})
    throughout the slab. In addition, cooling via 2-body interactions
    (e.g., bremsstrahlung) is enhanced, and
    the slab is able to efficiently cool even at large
    $h_f$. Therefore, the density of the heated layer appears to be
    crucial in developing a soft excess through a warm corona.\label{fig:nh5e14tau10spect}}
\end{figure*}

Figure~\ref{fig:nh5e14tau10spect} shows the results when the density
of the $\tau_{\mathrm{T}}=10$ layer has been increased by a factor of $2.5$.
The increase in density reduces the ionization parameter ($\xi=4\pi
F_{\mathrm{X}}/n_{\mathrm{H}}$, where $F_{\mathrm{X}}$ is the
flux of the illuminating power-law) of the slab. Thus, the inner few
Thomson depths are far less ionized by the X-ray power-law than the
lower density case seen in Fig.~\ref{fig:nh2e14tau10spect}. The impact
of the lower ionization states is readily seen in
Fig.~\ref{fig:nh5e14tau10spect} in the forest of X-ray recombination
lines at $\sim 1$~\kev\ and the weak neutral \fe\ line. The larger
density also significantly enhances the cooling rate by 2-body
processes, in particular bremsstrahlung and line cooling. Finally, the
coronal heating function $\mathcal{H} \propto n_{\mathrm{H}}^{-1}$
(Eq.~\ref{eq:hfunct}), so the heating rate provided by a fixed
$h_fD(R)$ is reduced for a larger density. These three effects all
work together so that it is very challenging to heat the
$n_{\mathrm{H}}=5\times 10^{14}$~cm$^{-3}$ slab close to the temperatures needed
to form a warm corona. The highest temperature reached (in the
$h_f=0.8$ model) is $\approx 0.3$~\kev, and Compton cooling never
dominates the cooling rates at any point in the slab. The emission and reflection
spectra shown in Fig.~\ref{fig:nh5e14tau10spect} show only a modest
soft excess forming at the highest $h_f$, with several spectral
features still remaining. It is apparent that the density at which the
coronal heating is injected will be critical in whether or not a warm
corona can develop.

\subsubsection{A Hydrostatic Atmosphere}
\label{subsub:hydro}
To further explore the effects of the density on the formation of a
warm corona, the calculations were repeated for an irradiated
atmosphere in hydrostatic balance\footnote{Note that this is a
  different experiment from the one performed by \citet{roz15}, as
  only the heated $\tau_{\mathrm{T}}=10$ layer is in hydrostatic
  balance. The connection to the cooler accretion disc underneath the
  layer is neglected.}. This set of models considers a
$\tau_{\mathrm{T}}=10$ layer at the surface of a radiation pressure
supported accretion disc $5$~Schwarzschild radii away from a $3\times
10^{7}$~M$_{\odot}$ black hole accreting at $0.05\times$ its Eddington
rate \citep{mfr00,brf01}. With these parameters $D(R)=3.594\times
10^{15}$~erg~cm$^{-2}$~s$^{-1}$. As in the constant density models,
the illuminating power-law flux is fixed at $0.1D(R)$ with
$\Gamma=1.9$ and $E_{\mathrm{cut}}=220$~\kev, and a coronal heating
flux $h_fD(R)$ is injected into the atmosphere with the heating
function $\mathcal{H}$ (Eq.~\ref{eq:hfunct}).

\begin{figure*}
  \includegraphics[width=0.98\textwidth]{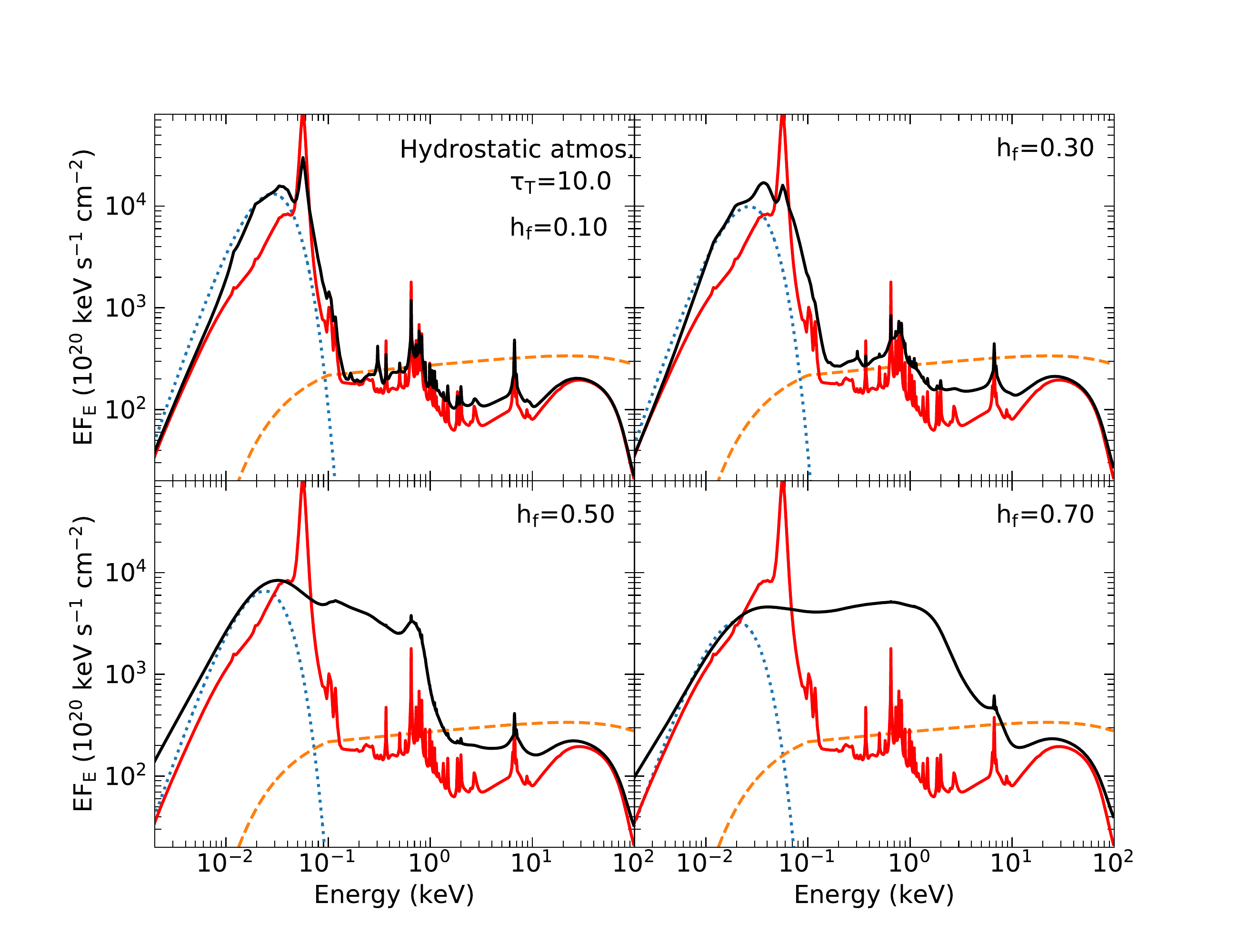}
  \caption{As in Fig.~\ref{fig:nh2e14tau10spect}, but now the
    $\tau=10$ layer is forced to be in hydrostatic equilibrium on the
    surface of a radiation pressure supported accretion disc. The
    hydrostatic atmosphere is subject to the same radiation conditions
  and heating processes as the constant density slabs, but with a
  smaller $D(R)$ (see text). As the atmosphere is in hydrostatic
  balance, the density falls off significantly with height
  \citep{nkk00,brf01}, reducing the efficiency of the two-body cooling
  processes, and allowing a strong soft excess to develop when $h_f
  \ga 0.5$.  \label{fig:hydrospect}}
\end{figure*}

Figure~\ref{fig:hydrospect} shows the emission and reflection spectra
produced by the heated and irradiated atmospheres.
The variation of spectral shape and properties with $h_f$ is similar
to the constant density case (Fig.~\ref{fig:nh2e14tau10spect}), with a
prominent soft excess forming at $h_f=0.5$ and a Comptonized
bremsstrahlung continuum seen when $h_f=0.7$. However, there are some
interesting differences in the spectra due to the density structure
adjusting to the heat deposited within it. For example, the overall
density of the atmosphere decreases as $h_f$ increases. When there is
no coronal heating the base density of the atmosphere is $4\times
10^{15}$~cm$^{-3}$, but the extra pressure caused by the heating
decreases the density throughout the atmosphere (e.g., the base
density is $3\times 10^{13}$~cm$^{-3}$ when $h_f=0.5$). This large-scale drop in
density will reduce the rates of 2-body processes throughout the
layer, increasing the temperature and ionization state of the gas, allowing the
formation of a warm corona. The lower density also reduces the
efficiency of photoelectric absorption, even at low values of $h_f$,
and thus the reflection spectra of the heated models is shifted
vertically upwards from the non-heated model in the upper two panels
of Fig.~\ref{fig:hydrospect}.

\begin{figure}
  \includegraphics[width=0.47\textwidth]{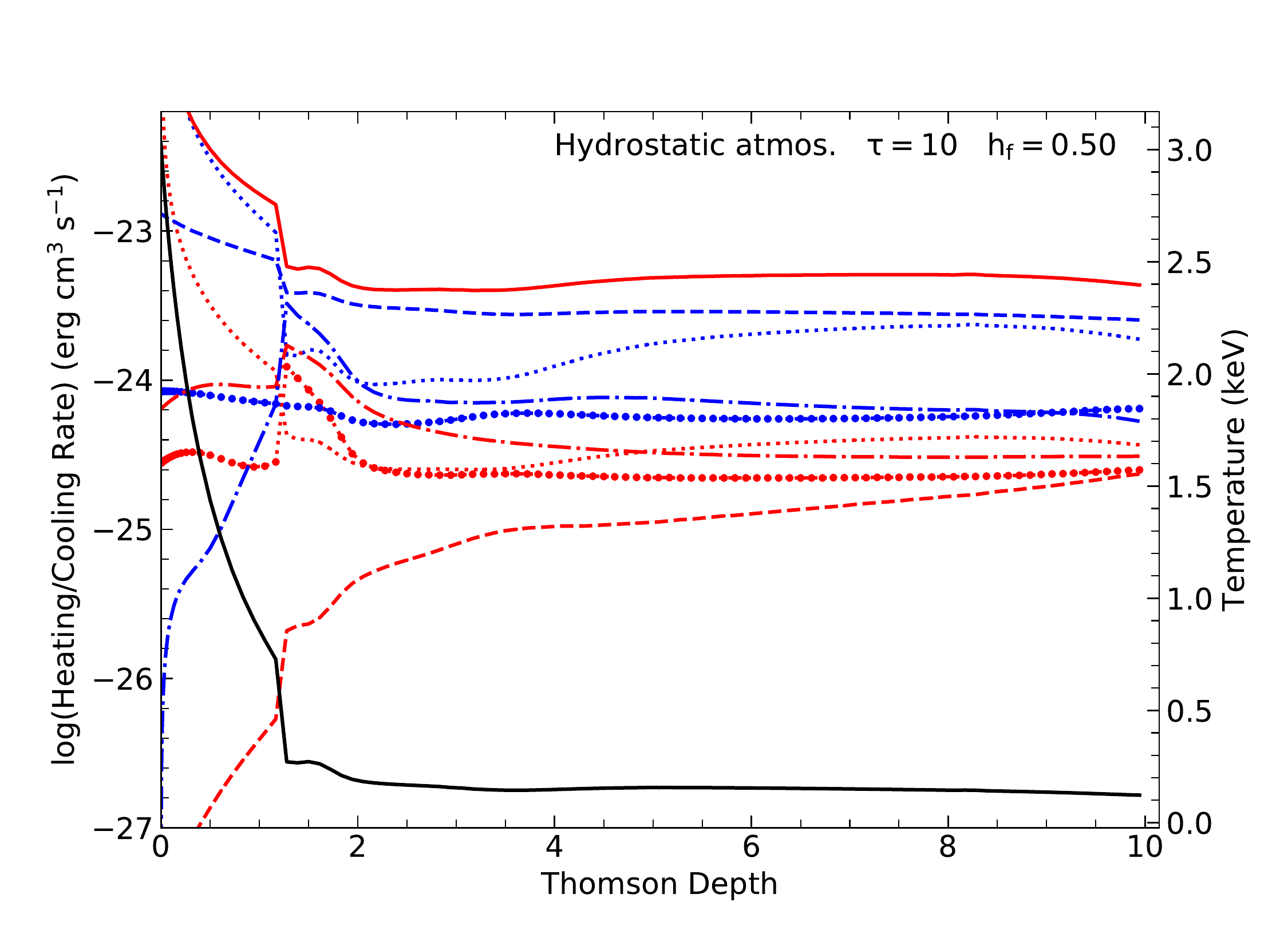}
  \caption{As in Fig.~\ref{fig:nh2e14rates}, but for a hydrostatic
    atmosphere with $h_f=0.5$. Since the coronal heating function
    $\mathcal{H} \propto n_{\mathrm{H}}^{-1}$ (Eq.~\ref{eq:hfunct}),
    the heating injected into the atmosphere increases strongly at
    $\tau_{\mathrm{T}} \la 1$ as the density drops with height. This
    strongly increases the temperature (sold black line), increasing the rate of Compton
    cooling (dotted blue line), leading to the soft excess in the
    emission spectrum (Fig.~\ref{fig:hydrospect}).  \label{fig:hydrorates}}
\end{figure}

The heating/cooling rates and temperature are plotted as a function
of $\tau_{\mathrm{T}}$ for the $h_f=0.5$ hydrostatic model in
Fig.~\ref{fig:hydrorates}. 
This figure clearly shows the impact of the variable density structure
on the properties of the layer. As $\mathcal{H} \propto
n_{\mathrm{H}}^{-1}$ (Eq.~\ref{eq:hfunct}), the heating of the gas
increases quickly as the density falls near the surface of the
atmosphere. This causes the temperature to rapidly increase to $\sim
1$--$2$~\kev. This warm, $\tau_{\mathrm{T}}\sim 1$ skin is dominated
by Compton cooling and will provide an effective scattering layer for
the thermal and line emission emerging from deeper in the
atmosphere. These experiments shows that a warm corona could develop
in the more realistic density structure of a hydrostatic atmosphere,
subject to the same conditions as the constant density slabs (i.e., as
long as the density is not too large, and the heating is not too
extreme). 

\section{Discussion}
\label{sect:discuss}

\subsection{Conditions Needed to Produce a Warm Corona and a Soft Excess}
\label{sub:conditions}
The results described above shows that accretion heat dissipated into a
$\tau_{\mathrm{T}}=10$--$20$ gas layer illuminated
by an X-ray power-law can, under certain conditions, lead to the
temperatures and spectral shapes consistent with a warm corona origin for
the AGN soft excess. However, as discussed below, the successful
warm corona models exist in a `Goldilocks' region of parameter space,
where the gas can neither be too hot nor too cold. Given the strong
density dependence on the heating and cooling rates, these temperature
requirements also translate into one on density. The exact values of
$n_{\mathrm{H}}$ and $h_f$ that lead to a warm corona will also depend on
$D(R)$. As a single representative value of $D(R)$ is assumed in this
investigation, the discussion on the dependence of the warm corona on
$n_{\mathrm{H}}$ and $h_f$ will necessarily be qualitative with the
understanding that if $D(R)$ increases over the value used here, the range of $h_f$
that leads to the formation of a warm corona will be shifted to
smaller values. Similarly, a larger $D(R)$ would allow a warm corona to exist at higher
densities than those considered in Sect.~\ref{sub:density}. 

Sect.~\ref{sub:tau} and~\ref{sub:density} showed that specific
conditions are needed in order for a warm corona with $kT \sim 1$~keV to
develop in the heated gas. If the coronal heating was too low than the
gas temperature would not rise high enough to fully ionise the metals
or for Comptonization to dominate the spectral shape below $\sim
1$~\kev. In these situations, the emitted spectrum would have a modest
soft excess overlayed with multiple recombination lines (e.g., top row
of Fig.~\ref{fig:nh2e14tau10spect}). From Eq.~\ref{eq:hfunct} it is
seen that such weak heating can occur if $h_f$ is too small, or if the
heated layer is too thick, as in Fig.~\ref{fig:nh2e14tau20spect}. In
the latter situation, where $\tau_{\mathrm{T}}=20$, a plausible warm
corona required the majority of the accretion flux $D(R)$ to be
released in the layer (e.g., Fig.~\ref{fig:nh2e14hf08}).

On the other hand, Fig.~\ref{fig:nh2e14tau10spect} also shows that if
$h_f$ grows too large, then the heated layer produces a Comptonized
bremsstrahlung spectrum that extends into the hard X-ray band. This
spectral shape could occur in rare circumstances (see
Sect.~\ref{sub:interesting}), but is unlikely to describe the shape of
the soft excess observed in the majority of AGNs. Therefore, if
dissipation of accretion energy in the upper layers of the accretion
disc is to be considered as a potential origin for the soft excess,
its strength must occur within a relatively narrow range so as to
provide enough heating to both significantly enhance the rate of Comptonization and
ionization, but is limited enough so as to not overheat the gas.

The density of the heated layer also plays a crucial role in the
ability of the warm corona to form in the heated layer. As both
line-cooling and bremsstrahlung are two body interactions, an increase
in density will boost the efficiency of both processes and thus
reducing the impact of the heating function $\mathcal{H}$. In addition, as
the heat was deposited on a per particle basis (Eq.~\ref{eq:hfunct}),
a fixed heat flux, $h_fD(R)$, will have less impact for larger
densities. Fig.~\ref{fig:nh5e14tau10spect} shows the outcome of the
constant density calculations when $n_{\mathrm{H}}=5\times
10^{14}$~cm$^{-3}$, a factor of $2.5$ larger than the ones shown in
Fig.~\ref{fig:nh2e14tau10spect}. Even with the largest $h_f$ the gas
temperature did not exceed $\sim 0.3$~\kev, and a smooth soft excess
never appeared in the spectrum. The experiments performed here strongly
suggest that a warm corona origin for the soft excess is likely to be
only viable in a heated layer that is not very dense. These conditions may be
naturally reached if the top $\tau_{\mathrm{T}} \sim 10$--$20$ layer of an
accretion disc falls off in density with height in a
manner similar to that needed for hydrostatic
balance. Fig.~\ref{fig:hydrospect} showed that this scenario can
produce appropriate soft excesses and warm coronae with $kT \sim 1$~keV.

The role of the external power-law is also found to be crucial in
the formation of a warm corona. The X-ray power-law provides a base
level of heat and ionization throughout the gas layer, reducing the
atomic opacity at all energies. Thus, the emitted spectra (e.g., Fig.~\ref{fig:nh2e14tau10spect}), which is
the combination of the transmitted blackbody, thermal emission from
the gas, and scattered power-law photons, do not exhibit strong absorption
lines below $\sim 1$~\kev, as found by \citet{garcia19}. We verified
with our constant density models that the absorption lines appeared
upon removing the power-law from the calculation. Thus, if a warm
corona exists in AGN accretion discs, it must occur along with the hot
corona that produces the hard X-ray power-law. Such a scenario would
provide constraints on the vertical transport of accretion energy
through the disc.

The predicted emission and reflection spectra show that the impact of
a $kT \sim 1$~\kev\ warm corona has additional effects on the X-ray
spectra beyond producing a smooth soft excess. The extra heating
provided by $\mathcal{H}$ increases the ionization state of the gas
and thus an ionised \fe\ line at $6.7$~\kev\ is prominent in all the
output spectra. Ionised \fe\ lines are not commonly observed in most
Seyfert galaxies, including those with soft excesses \citep[e.g.,][]{walton13}. This may
indicate that a warm corona is not the correct model for producing the
AGN soft excess. However, the models calculated here considered only a
single radius and one value of $D(R)$, while X-ray observations probe the integrated emission
from the inner accretion disc, as well as detecting emission
reprocessed from much larger distances \citep[e.g.,][]{patrick12}. Therefore, a more
comprehensive investigation of the warm corona scenario, which
accounts for the radial dependence of $D(R)$ and (potentially) $h_f$,
is needed to better compare to observations. The development of such a
model, including the ability to fit AGN X-ray spectra, is planned for
future work.

Given that it is possible that both a warm and a hot corona exist in
the centers of accretion flows, it appears likely that the soft excess
is actually produced by a combination of both Comptonized thermal
emission and reprocessed hard X-ray emission. The multi-epoch spectral modeling of \mrk\ by \citet{kb16}
showed that, upon subtracting the emission predicted from reflection,
the remaining soft excess flux tracked the overall flux level of the
source, indicating that it was influenced by the underlying accretion
physics. The soft
excesses presented in Sect.~\ref{sect:results} are comprised of both
emission from the Comptonized blackbody and emission from gas heated
by the power-law.  We have neglected the influence of relativistic blurring for
the purposes of this paper, but if this emission is emerging from
close to the inner edge of the accretion disc, then relativistic
smearing will further smooth out the predicted soft excesses. This
effect will allow even relatively small values of $h_f$ to produce
smooth soft excesses. The combined effects of both the hard and warm corona,
and how they change in strength with radius and accretion rate, may
lead to the variety of spectral features and shapes observed in AGNs.

\subsection{A Warm Corona Connection to Narrow-Line Seyfert 1 Galaxies?}
\label{sub:interesting}
The results of the warm corona calculations showed that the excess
heating in a $\tau_{\mathrm{T}}=10$--$20$ gas layer can produce an interesting
range of potential spectral shapes. The most unusual ones were the
$h_f=0.7$ model shown in Figs.~\ref{fig:nh2e14tau10spect}
and~\ref{fig:hydrospect} which are dominated by a Comptonized
bremsstrahlung continuum that leads to a steep spectrum at energies
$\ga 2$~\kev. It is interesting to consider if such unusual spectra
may be related to ones observed in AGNs, in particular Narrow-Line
Seyfert 1 galaxies (NLS1s; \citealt{gallo18}).

\begin{figure}
  \includegraphics[width=0.47\textwidth]{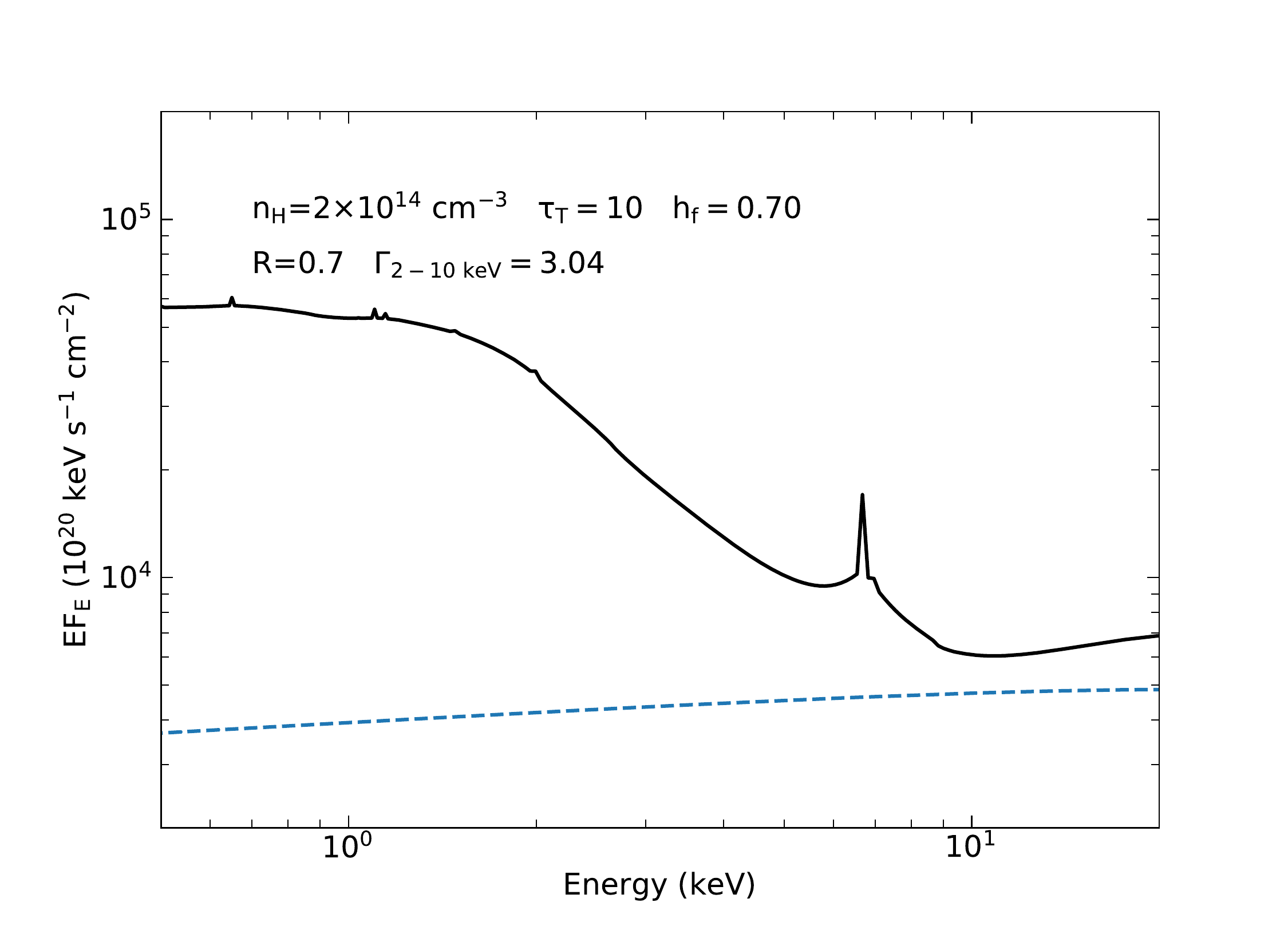}
  \caption{The black line is the emission and reflection spectrum from
  the $n_{\mathrm{H}}=2\times 10^{14}$~cm$^{-3}$,
  $\tau_{\mathrm{T}}$=10, $h_f=0.70$ model upon adding it to the
  illuminating power-law (dashed line) so that it has a reflection
  fraction of $R=0.7$. The $2$--$10$~\kev\ photon-index of the
  resulting spectrum is $3.04$. The warm corona scenario appears to
  provide a mechanism with which to produce the steep AGN spectra and
  ionised \fe\ lines seen in NLS1s \citep[e.g.,][]{gallo18}.  \label{fig:spect}}
\end{figure}

Figure~\ref{fig:spect} plots the $h_f=0.7$ spectrum from the
$n_{\mathrm{H}}=2\times 10^{14}$~cm$^{-3}$, $\tau_{\mathrm{T}}=10$
series of models, but we have now added the illuminating power-law to
the reflection spectrum, so that $R$, the reflection fraction, is
$0.7$, consistent with typical values observed in AGNs
\citep[e.g.,][]{zappa18}.
The spectrum shown in the figure has a steep hard X-ray spectrum with
a $2$--$10$~\kev\ photon-index of $3.04$, an ionised \fe\ line and
deep iron edge, all properties similar to those observed in several NLS1s (e.g., \oneh;
\citealt{bif01,boll02,zog10,fab12,liu16}). Many NLS1s are consistent with being relatively low mass supermassive
black holes that are accreting at a high fraction of the Eddington
limit \citep[e.g.,][]{williams18}, suggesting that the steep X-ray
spectra are a consequence of cooling of the hot corona due to the intense
radiation field produced by the rapidly accreting disc. Alternatively,
Fig.~\ref{fig:spect} suggests that heating of the accretion disc
surface and the formation of a warm corona can also explain the
interesting features of NLS1s. As the accretion rate increases, the disc is expected to
grow thicker, perhaps enveloping some of the hot corona into the
surface of the disc. Thus, a large fraction of the overall accretion
energy is dissipated within the surface of the disc producing a strong
warm corona that would lead to this steep spectrum. In this scenario,
less luminous AGNs, home to less rapidly accreting black holes, would
have thinner discs, and therefore smaller $h_f$ at their surface
yielding moderate warm corona that, along with the reflection
spectrum, produce the soft excesses observed in typical AGNs. It
appears plausible that if accretion energy is dissipated into the
surface layers of the disc, $h_f$ would vary as the disc changes in height
and density structure, leading to spectral shapes that directly track
these effects. 

\section{Conclusions}
\label{sect:concl}
The origin of the soft excess observed in the majority of AGN X-ray
spectra has remained poorly understood for over three decades. One
compelling potential explanation has been that a warm corona is generated at the
surface of the accretion disc which Compton scatters thermal emission
from the disc into the broad, largely featureless spectrum of the soft
excess. A phenomenological model describing this idea can fit the
spectra of multiple AGNs and indicates that a temperature of $\sim
1$~\kev\ and an optical depth of $\tau_{\mathrm{T}}\approx 10$--$20$
is needed for a successful warm corona 
\citep[e.g.,][]{petrucci18}, raising potential theoretical problems
with this scenario \citep{roz15,garcia19}.

This paper considered the physical conditions and emitted spectra of a
$\tau_{T}=10$ or $20$ gas layer subject to irradiation from a X-ray
power-law from above, a blackbody from below, and has a variable
amount of excess internal heat (proportional to the accretion flux
$D(R)$). We found that warm corona temperatures and strong,
featureless soft excesses could be produced in some models, but only for
a particularly narrow range of heating conditions and gas
densities. If the internal heating was not sufficient the gas remained
too cool and Compton scattering could not produce a strong soft
excess; alternatively, if too large a fraction of $D(R)$ is dissipated
in the layer, the gas reaches so high a temperature that the emission
spectrum is dominated by a Comptonized bremsstrahlung
continuum. Similarly, it is found that the density of the gas can
not be too large since the gas would be able to efficiently cool through 2-body
processes, inhibiting the formation of a warm corona. This density condition may be
alleviated if the density of the top $\tau_{T} \approx 10$ layer falls
with height as if in hydrostatic balance. Finally, the hard X-ray
power-law was found to be crucial in producing a warm corona, as it
raised the ionization state of the surface high enough to avoid
imprinting the emitted spectrum with strong absorption lines.

Although the conditions for formation are relatively narrow, we
conclude that a warm corona is a viable scenario for contributing to
the soft excess in AGNs.
The spectra calculated here have
soft excesses that are produced from the combination of Comptonized
disc emission and gas heated by the X-ray power-law. As it is possible
for both warm and
hot corona to co-exist in AGNs, then the soft excess is likely a product of
both origins working together.

Future work on this model will focus on adding the radial dependence
of $D(R)$ and generating a model that can be fit to X-ray spectra of
AGNs to determine the relative importance of warm and hot corona. In
particular, as the properties of the spectra change so dramatically as
$h_f$ changes, there is the tantalizing possibility of correlating
these effects with fundamental AGN parameters such as luminosity and
accretion rate. Measuring such changes would lead to important
constraints on how accretion energy is transported and dissipated
in accretion discs.

\section*{Acknowledgements}
The author thanks A.\ Fabian for useful discussions.




\bibliographystyle{mnras}
\bibliography{refs} 




\bsp	
\label{lastpage}
\end{document}